\documentclass[a4paper]{article}

\usepackage{icrc2013}
\usepackage[numbers]{natbib}
\usepackage{overpic}

\usepackage[english]{babel}

\title{The renaissance of radio detection of cosmic rays}

\shorttitle{Tim Huege --- The renaissance of radio detection of cosmic rays}

\authors{Tim Huege$^{1}$}
\afiliations{$^1$IKP, Karlsruhe Institute of Technology (KIT), Postfach 3640, 76021 Karlsruhe, Germany}
\email{tim.huege@kit.edu}

\abstract{Nearly 50 years ago, the first radio signals from cosmic ray air showers were detected. After many successful studies, however, research ceased not even 10 years later. Only a decade ago, the field was revived with the application of powerful digital signal processing techniques. Since then, the detection technique has matured, and we are now in a phase of transition from small-scale experiments accessing energies below 10$^{18}$ eV to experiments with a reach for energies beyond 10$^{19}$~eV. We have demonstrated that air shower radio signals carry information on both the energy and the mass of the primary particle, and current experiments are in the process of quantifying the precision with which this information can be accessed. All of this rests on a solid understanding of the radio emission processes which can be interpreted as a coherent superposition of geomagnetic emission, Askaryan charge-excess radiation, and Cherenkov-like coherence effects arising in the density gradient of the atmosphere. In this article, I highlight the ``state of the art'' of radio detection of cosmic rays and briefly discuss its perspectives for the next few years.}

\keywords{radio detection, cosmic rays}

\begin{document}
\maketitle

\section{Introduction}

One hundred years after the discovery of cosmic rays, their sources are still a mystery --- in particular at the highest energies. To make significant progress, we need to measure the arrival direction, the energy, and the mass of individual cosmic rays. Due to the low flux at the highest energies, acquiring acceptable statistics requires very large experiments running with high duty cycles. Currently established techniques have been very successful, but they still have limitations: particle detectors only sample a tiny snapshot of the shower evolution and suffer from uncertainties in the hadronic interaction models. Fluorescence detectors provide us with a calorimetric energy measurement and allow us to directly determine the depth of the air shower maximum, a mass-sensitive parameter, but they achieve duty cycles of only $\approx 10$\%.

Radio detection of cosmic rays sets out to provide information which is highly complementary to that gathered by existing techniques, in particular particle detectors. Radio detection promises to directly probe the electromagnetic component of the air shower, provide us with very good angular resolution, a calorimetric energy measurement, sensitivity to $X_{\mathrm{max}}$, and a duty cycle of $\approx 100$\%. In addition, radio detectors can potentially be built cheaply. But how well does all of this work in practice?

As I will discuss in the following, radio detection has left the stage of small prototype experiments and is now making the step to large-scale application. The radio emission mechanisms are understood, and we are on the verge of establishing the precision achievable with radio measurements.


\section{The renaissance of radio detection}

Radio detection of cosmic rays is not a new technique. The initial detection of pulsed radio emission being emitted by air showers was made by Jelley et al. in 1965 \citep{JelleyFruinPorter1965}. A period of intense research followed, but the activities stopped completely in the 1970s. A lot was already clear at that stage (see in particular the review article of Allan \citep{Allan1971}): the radio emission is of dominantly geomagnetic origin, the radio lateral distribution falls off roughly exponentially, signals were detected from 2 to 500 MHz,  and the electric field amplitude grows approximately linearly with energy. However, there were also many open questions and issues where different groups came to different conclusions: How large are the absolute field strenghts? Are there emission mechanisms in addition to the geomagnetic one? Do atmospheric electric fields destroy the correlations of signal strength and energy of the cosmic ray primary? And can radio measurements be used to extract information on $X_{\mathrm{max}}$?

\begin{figure*}[!htb]
  \vspace{2mm}
  \centering
  \includegraphics[width=0.72\textwidth]{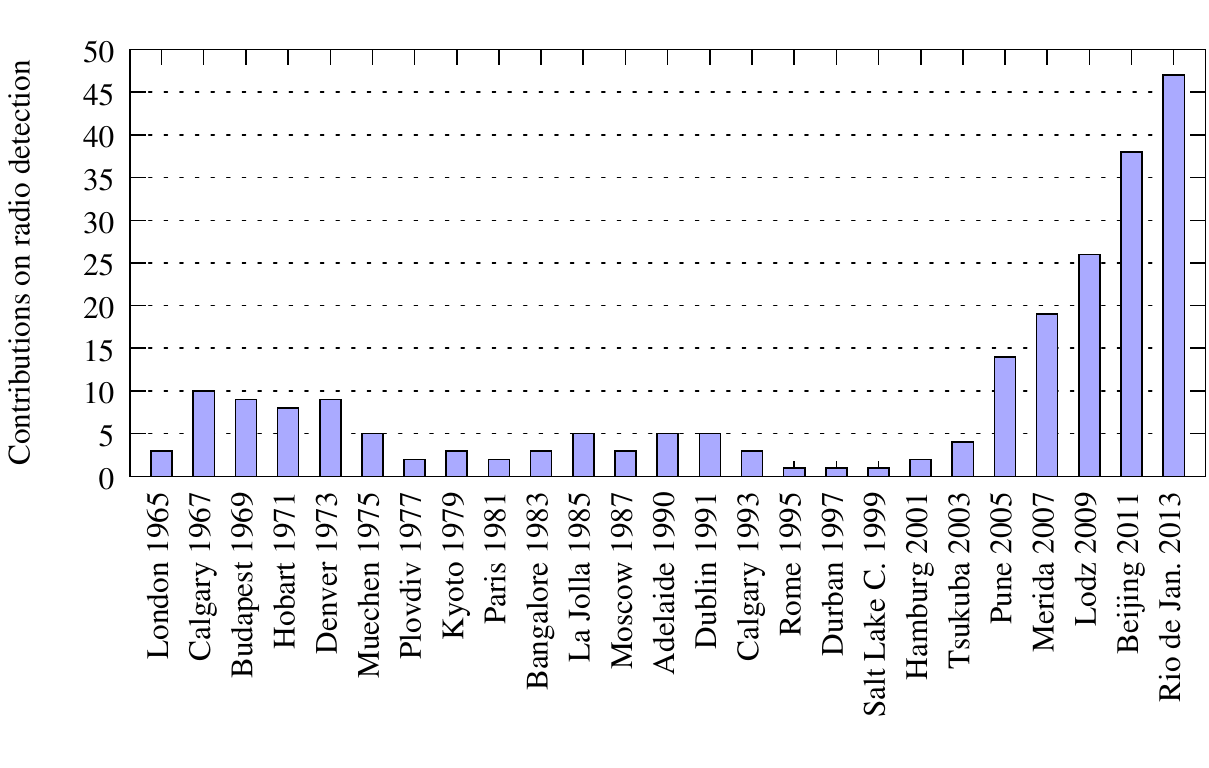}
  \vspace{-7mm}
  \caption{Number of contributions related to radio detection of cosmic rays or neutrinos to the ICRCs since 1965. The field has grown very impressively since the modern activities started around 2003. Data up to 2007 were taken from \citep{Nichol2011}.}
  \label{fig:icrcstats}
 \end{figure*}

With the advent of modern digital signal processing techniques, radio detection of cosmic rays came into focus again roughly a decade ago, with the start of the LOPES \citep{FalckeNature2005} and CODALEMA \citep{ArdouinBelletoileCharrier2005} experiments measuring in the frequency range between the AM band at $\sim 20$~MHz and the FM band at $\sim 80$~MHz. Both experiments have access to energies of up to $\sim 10^{18}$~eV. They started as small prototype installations and have been re-configured, extended and improved many times. A number of other small-scale projects followed: the TREND experiment \citep{Trend2011}, prototype detectors at the Pierre Auger Observatory \citep{RevenuRAuger2012} and radio detectors at the YAKUTSK array \citep{PetrovYakutsk2013}. The great successes of these first-generation modern MHz experiments prompted plans to apply the technique on much larger scales in a set of second-generation modern MHz experiments: the Auger Engineering Radio Array (AERA) \citep{SchroederAERAIcrc2013}, aiming at energies of up to $> 10^{19}$~eV, the Tunka radio extension \citep{SchroederTRexIcrc2013} (Tunka-Rex) and the Low Frequency Array \citep{NellesIcrc2013} (LOFAR). In only 10 years, the field has made very good progress and achieved a high degree of maturity. Radio detection of cosmic rays and neutrinos has indeed grown strongly in these last few years, as is illustrated impressively by the number of constributions to the International Cosmic Ray Conferences displayed in Fig.\ \ref{fig:icrcstats}.

\subsection{Radio emission physics}

When LOPES and CODALEMA started, we did not have a good understanding of the radio emission physics. This has changed dramatically in the past few years. Numerous modelling approaches have been developed, from macroscopic ones describing the emission physics using currents and net charge in the air shower (e.g. MGMR \citep{DeVriesVanDenBergScholten2010}, EVA \citep{WernerDeVriesScholten2012}) over models based on histogrammed particle distributions or air shower universality (e.g. REAS3.1 \citep{LudwigHuege2010} and SELFAS2 \citep{MarinRevenu2012}) to microscopic full Monte Carlo simulations following the radiation emitted due to the acceleration of individual electrons and positrons in the air shower (e.g. CoREAS \citep{HuegeARENA2012a} and ZHAireS \citep{AlvarezMunizCarvalhoZas2012}). A detailed discussion of the models is beyond the scope of this article, please refer to \citep{HuegeARENA2012b} for further information. The most important point is, however, that all of these models deliver comparable results. A consistent picture has emerged, which I will discuss in some more detail.

\subsubsection{Emission contributions}

 \begin{figure*}[!htb]
  \vspace{2mm}
  \centering
  \includegraphics[width=0.24\textwidth]{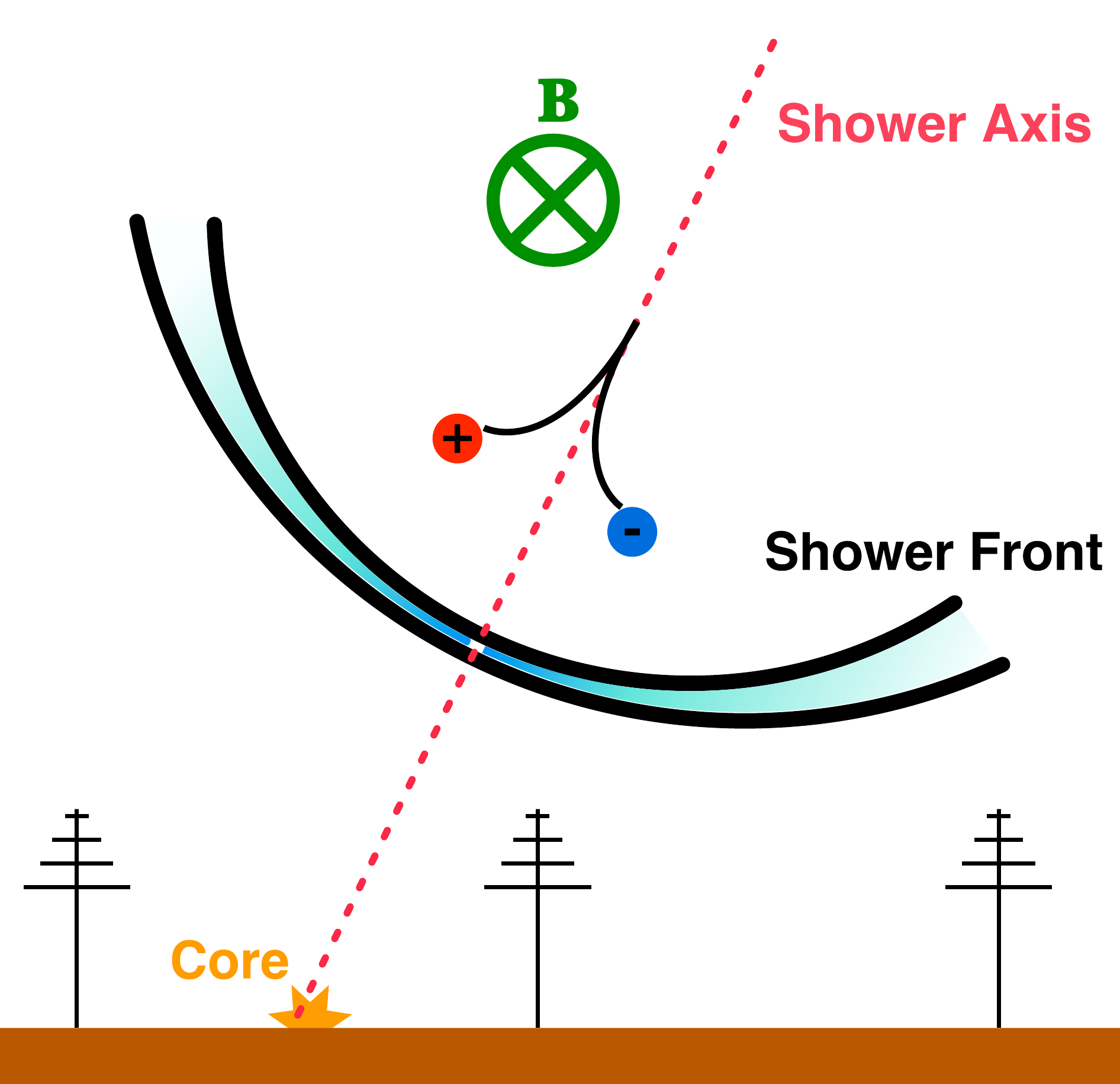}
  \includegraphics[width=0.22\textwidth]{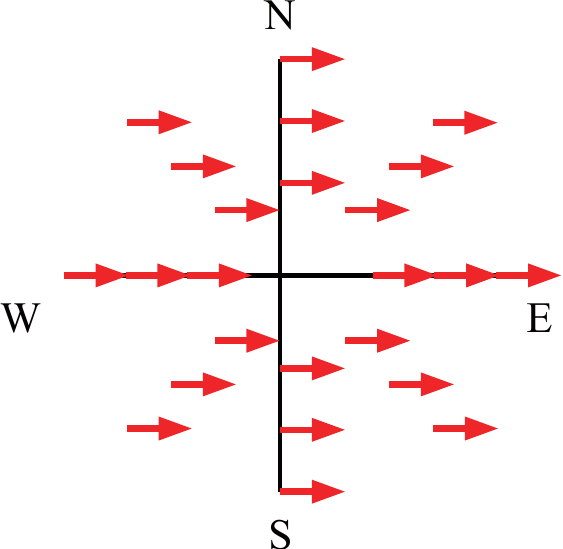}
  \hspace{0.06\textwidth}
  \includegraphics[width=0.24\textwidth]{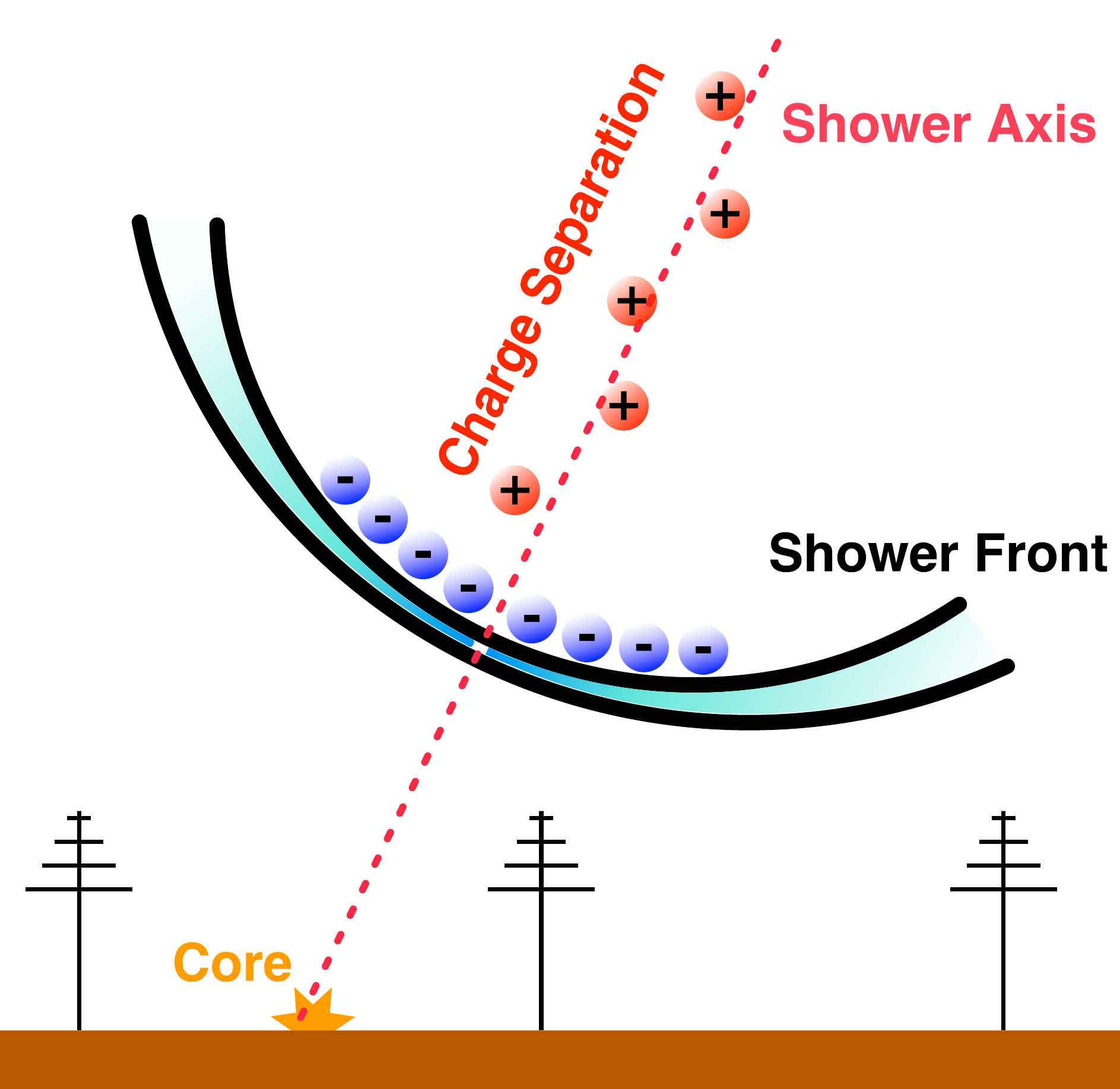}
  \includegraphics[width=0.22\textwidth]{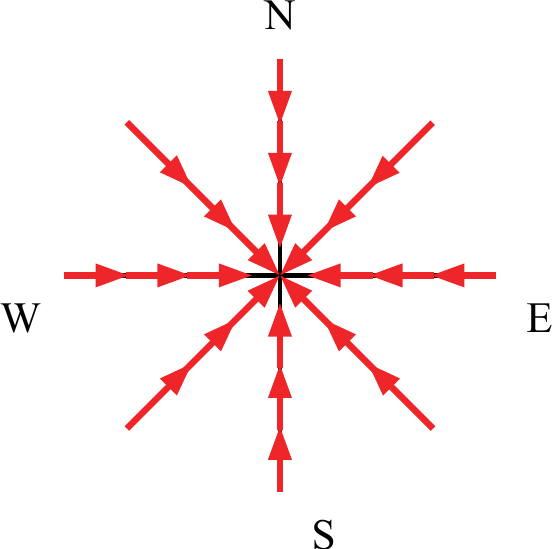}
  \caption{Schematic representation of the two main mechanisms contributing to the radio signal from extensive air showers. Time-varying transverse currents initiated by the geomagnetic field constitute the primary effect. The emission is linearly polarized in the direction given by the Lorentz force, irrespective of observer position (left). A time-varying negative charge excess constitutes the secondary effect. This is the Askaryan effect, which is the main emission mechanism in dense media, but is sub-dominant in air showers. The emission is linearly polarized, with the electric field vector being aligned radially with respect to the shower axis (right). Diagrams are from \citep{SchoorlemmerThesis2012} and K.D. de Vries.}
  \label{fig:mechanisms}
 \end{figure*}

\begin{figure}[!htb]
  \vspace{2mm}
  \centering
  \includegraphics[width=0.49\textwidth]{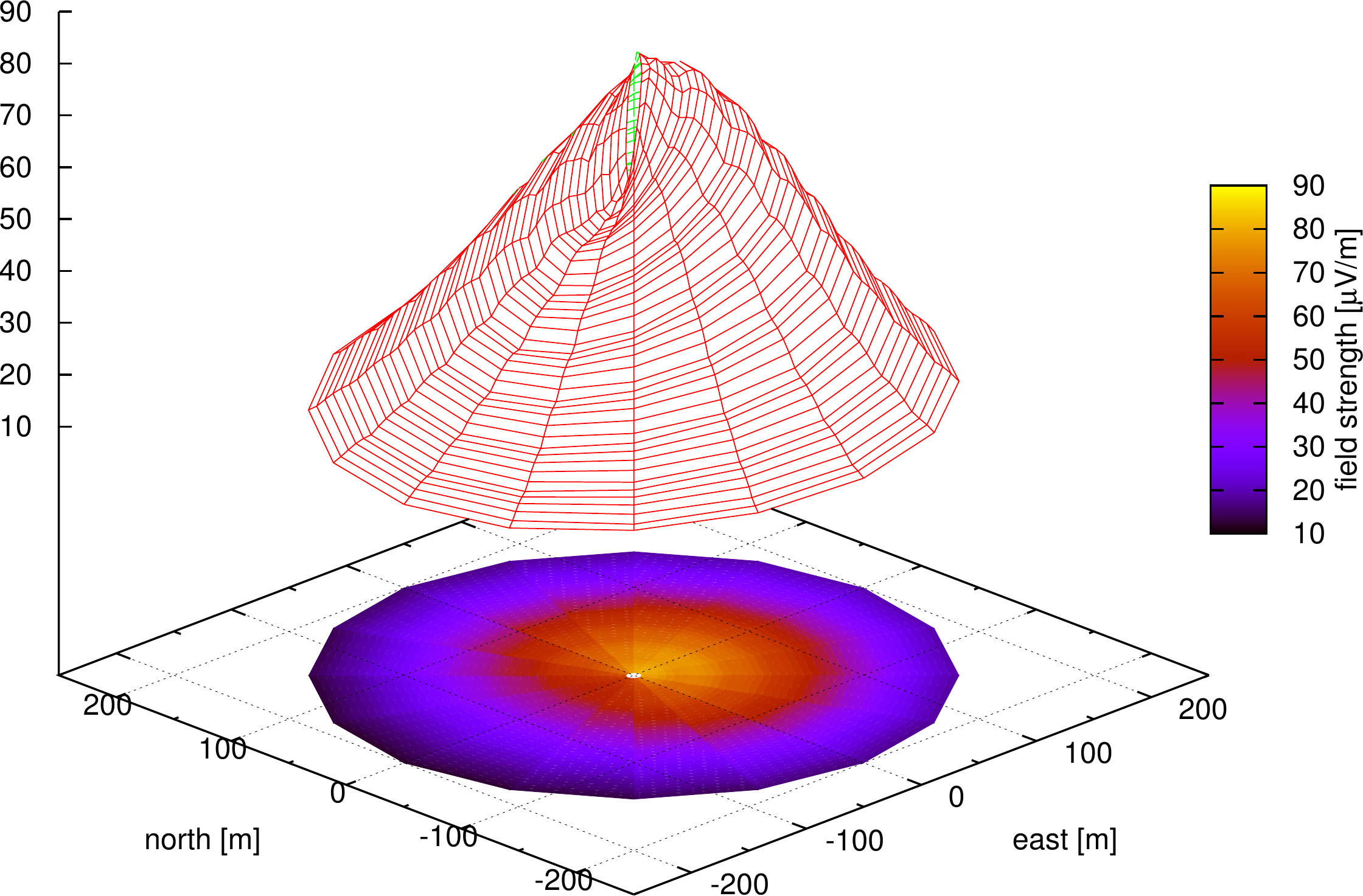}
  \caption{The superposition of the geomagnetic and charge-excess emission leads to an asymmetric, two-dimensional lateral distribution of the radio signal, as shown here for a CoREAS simulation \citep{HuegeARENA2012a}. If measured in detail, this reveals a wealth of information on the air shower properties.}
  \label{fig:footprint}
 \end{figure}

Two main emission mechanisms contribute to the radio signal. The dominating contribution is of geomagnetic origin. Electrons and positrons in the air shower are accelerated in the geomagnetic field. At the same time, they are decelerated in interactions with air molecules. This leads to a net drift of electrons and positrons and thus to a current, similar to the situation in a conductor. These transverse currents grow as the air shower grows and decline when the air shower dies out after reaching its maximum. These {\em time-varying transverse currents} are the primary source of radiation. The radiation is linearly polarized and the electric field vector is oriented in the direction of the Lorentz force, irrespective of the position of the observer, as illustrated in Fig.\ \ref{fig:mechanisms} (left). The second mechanism is the Askaryan-effect \citep{Askaryan1962a,Askaryan1965}, which is also responsible for the radio emission of showers in dense media\footnote{It should be stressed that Askaryan emission is not ``classical Cherenkov emission'', please see \citep{JamesFalckeHuege2012} for details.}. An excess of electrons over positrons is present in the air shower, and again, the shower evolves and thus leads to a {\em time-varying net charge} which provides the secondary contribution to the radio signal. The charge-excess emission is also linearly polarized, but the electric field vector is oriented radially with respect to the shower axis, i.e., its orientation depends on the position of the observer, as illustrated in Fig.\ \ref{fig:mechanisms} (right). Superposition of these two contributions leads to a complex lateral distribution of the radio signal exhibiting prominent asymmetries, as is illustrated in Fig.\ \ref{fig:footprint}.

\subsubsection{Evidence for charge-excess emission}

 \begin{figure*}[!htb]
  \vspace{2mm}
  \centering
  \includegraphics[width=0.29\textwidth]{icrc2013-1294-07.pdf}
  \includegraphics[width=0.36\textwidth]{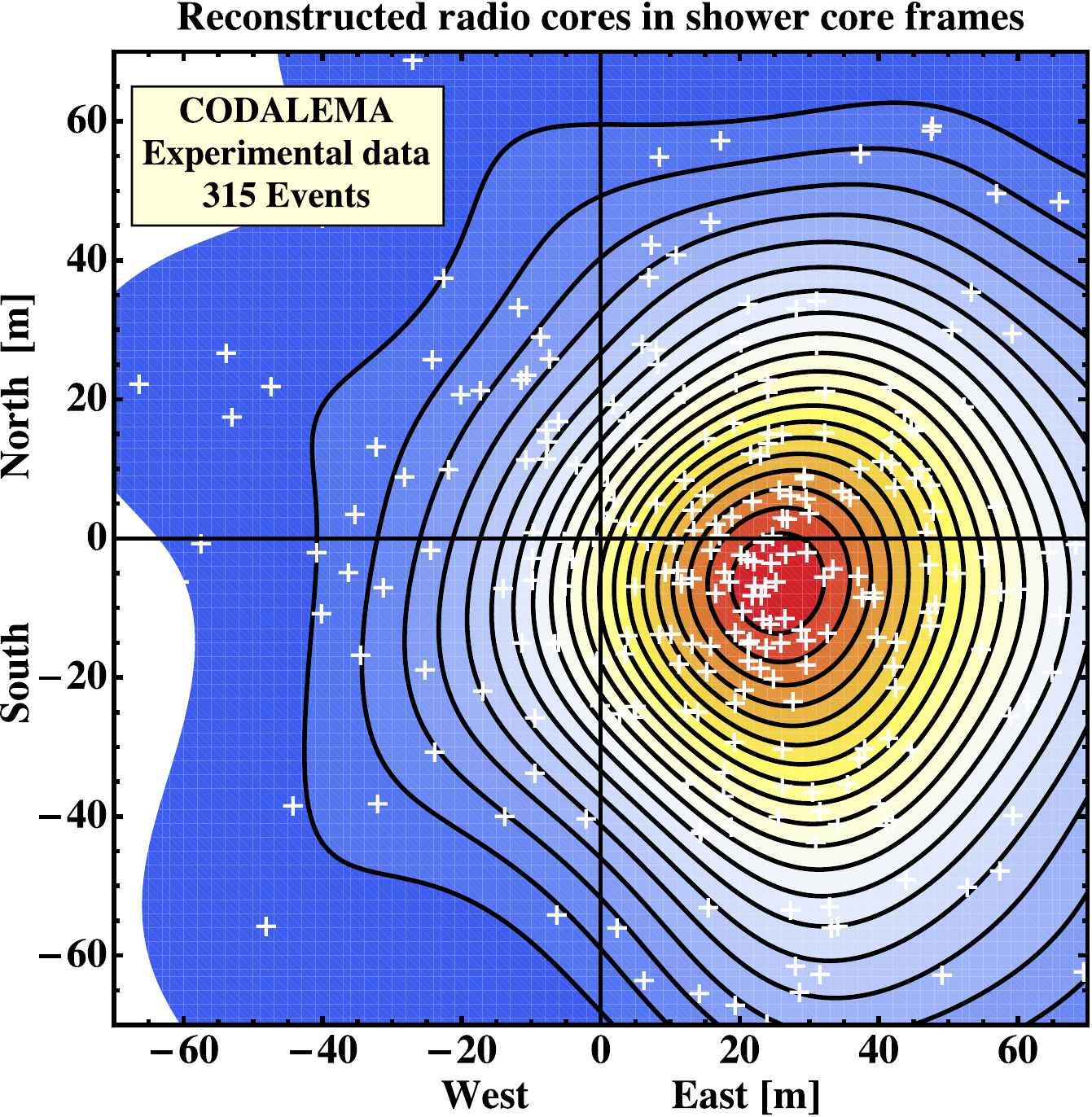}
  \includegraphics[width=0.27\textwidth]{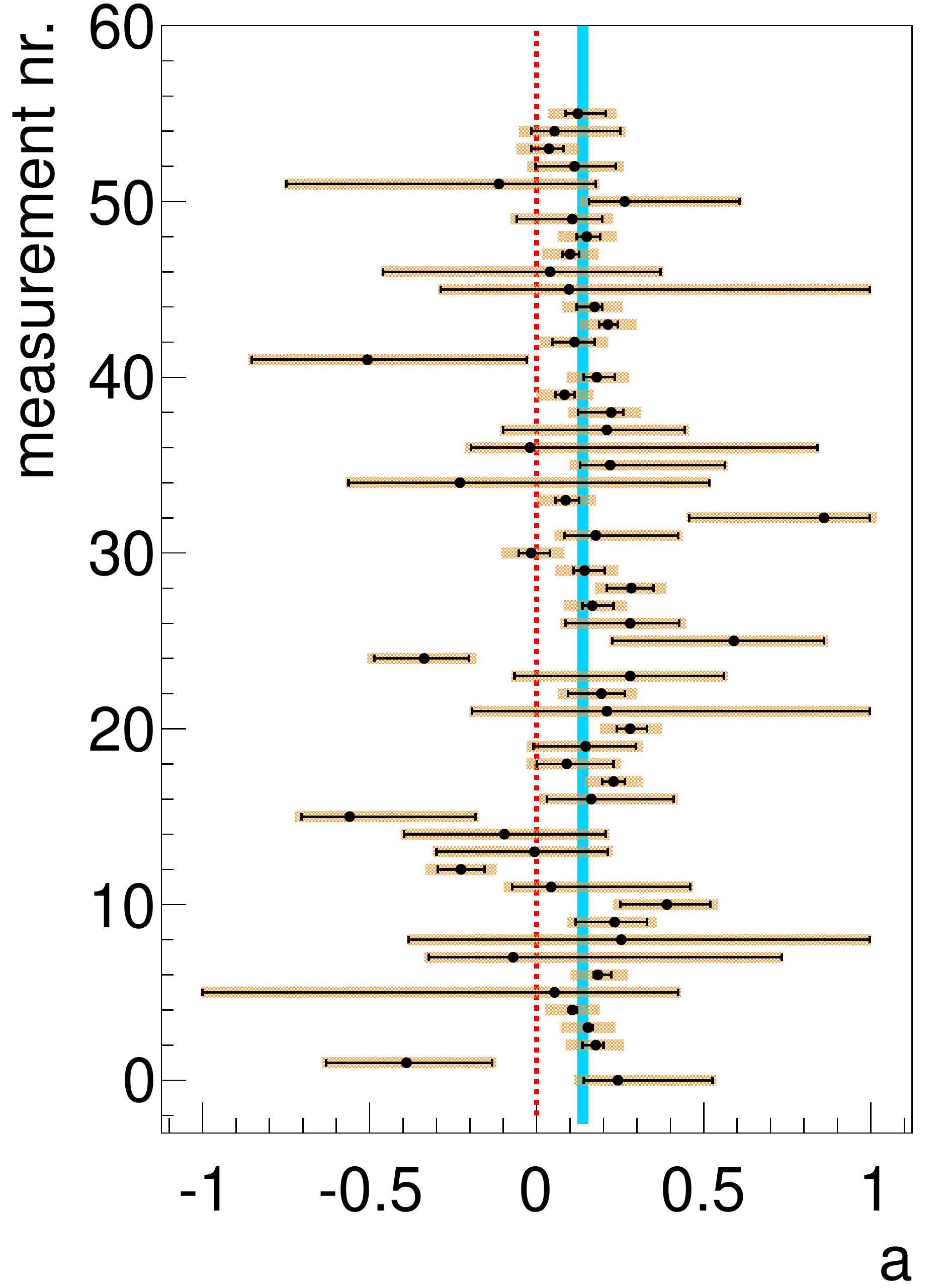}
  \caption{The presence of the sub-dominant charge excess contribution to the EAS radio emission has been shown by Prescott et al.\ \citep{PrescottHoughPidcock1971} (left, 14\% signal remaining for showers parallel to the geomagnetic field), CODALEMA \citep{CODALEMACoreShift} (middle, shift of the core positions reconstructed from radio data due to the asymmetry caused by the superposition of geomagnetic and charge-excess radiation) and AERA \citep{HuegeAERAIcrc2013} (right, presence of a radially oriented electric field contribution with an average strength of 14\%).}
  \label{fig:chargeexcess}
 \end{figure*}

The dominance of geomagnetic emission was already well-established in the 1970s. The charge-excess contribution was predicted theoretically by Askaryan, but generally there was no clear indication that it was needed to explain the measurements. In 1971, one analysis \citep{PrescottHoughPidcock1971} showed that the radio emission did not vanish even for showers oriented parallel to the local geomagnetic field as is visible in Fig.\ \ref{fig:chargeexcess} (left). This was an indication of a contribution in addition to the geomagnetic one at the $\sim 14$\% level which was presumed to be the charge-excess effect, but in fact no further information on the properties of this contribution was available. A clear and unambiguous identification was achieved only recently with the modern experiments. An analysis of CODALEMA \citep{CODALEMACoreShift}, shown in Fig.\ \ref{fig:chargeexcess} (middle) found a systematic shift of the core positions reconstructed from radio data with respect to the core positions reconstructed from particle detectors. This is consistent with the asymmetries in the radio signal caused by the superposition of geomagnetic and charge-excess emission (Fig.\ \ref{fig:footprint}). Even more direct proof was recently provided by AERA \citep{HuegeAERAIcrc2013}, which measured the orientation of the measured polarization vector as a function of the antenna location with respect to the shower axis and found that the data are not explainable by pure geomagnetic radiation. Adding a linearly polarized contribution with radial orientation of the electric field vector --- as expected for the charge-excess emission ---, however, the measurements can be explained. In Fig.\ \ref{fig:chargeexcess} (right) the relative strength $a$ of the necessary radial component with respect to the geomagnetic component is displayed for each individual measurement (antenna). While the value of $a$ shows significant scatter, the source of which is currently not fully understood, an average contribution of 14\% of an emission component with radial electric field orientation was found. 

\subsubsection{Coherence and Cherenkov effects}

Two further aspects play an important role in the emission physics. At MHz frequencies, the two mechanisms radiate coherently, i.e. the electric fields (rather than the intensities) radiated by individual particles add up. The reason is that the source region is smaller than the wavelength at these frequencies.

Secondly, the air shower front moves approximately with the speed of light, whereas the radio emission propagates at slightly lower velocity due to the density-dependent refractive index of the atmosphere. For observers along specific angles to the shower axis, this leads to Cherenkov-like time compression of the radio emission produced in the above-described mechanisms \citep{DeVriesBergScholten2011}. This time-compression leads to very short pulses which can have significant power at high frequencies. Indeed, there are now experimental results by the CROME experiment \citep{SmidaCROMEIcrc2013b} that air shower radio signals observed at frequencies of 3.4 to 4.2~GHz can be reasonably explained with this picture. Also, it is very likely that the cosmic ray events detected by ANITA in the frequency range of 300 to 1200 MHz \citep{HooverNamGorham2010} were such Cherenkov-boosted radio signals from cosmic ray air showers.

 \begin{figure}[!htb]
  \vspace{2mm}
  \centering
  \includegraphics[width=0.48\textwidth]{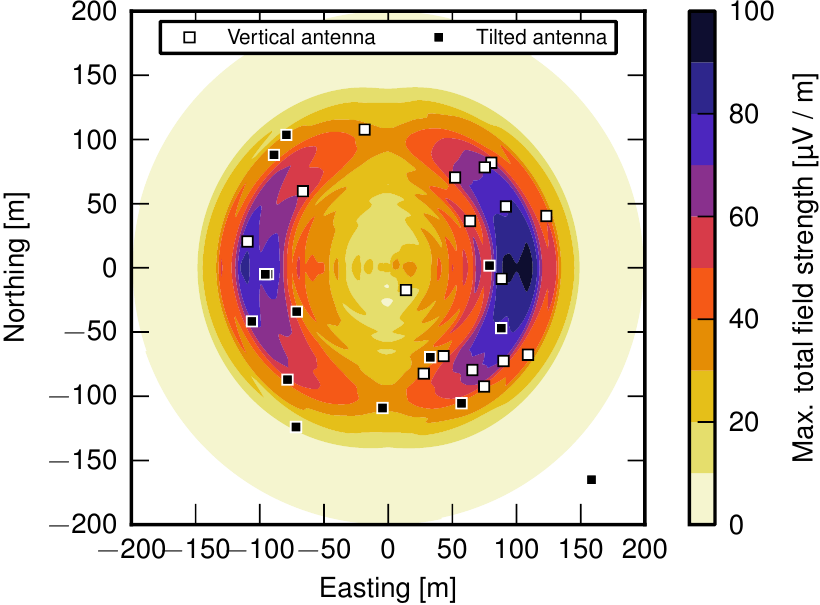}
  \caption{Radio emission from extensive air showers has been detected at frequencies of 3.4--4.2~GHz by the CROME experiment \citep{SmidaCROMEIcrc2013b}. The distribution of the core positions of detected showers is concentrated on a Cherenkov ring, as expected from CoREAS simulations overlaid in color scale.}
  \label{fig:crome}
 \end{figure}

\subsubsection{Quality of the modelling}

\begin{figure*}[!htb]
  \vspace{2mm}
  \centering
  \includegraphics[width=0.9\textwidth]{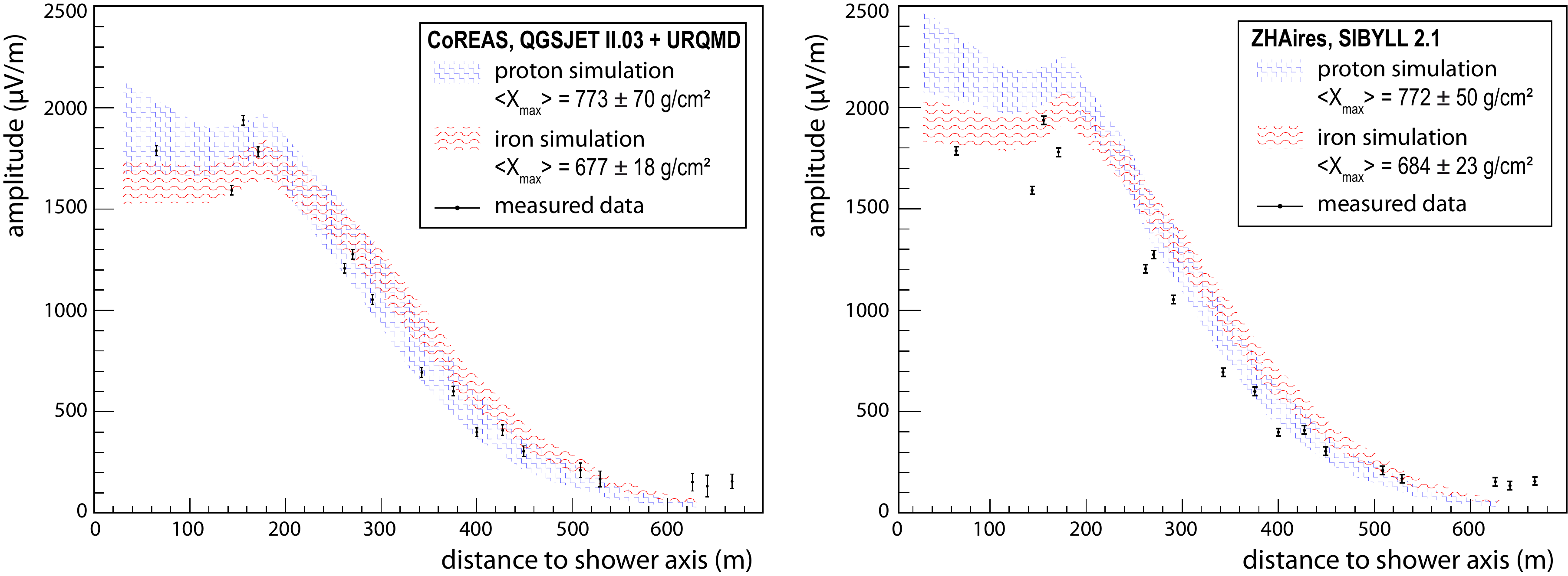}
  \caption{Comparison of the electric field amplitudes measured by AERA for a given radio event with simulations of CoREAS and ZHAireS \citep{SchroederAERAIcrc2013}. The simulations include shower-to-shower fluctuations, the uncertainties of the simulation input parameters taken from a reconstruction with the Auger infill surface detector array, and a complete detector simulation of AERA. The agreement is very good, including the absolute scale of the emission.}
  \label{fig:aeracomparison}
 \end{figure*}

\begin{figure}[!htb]
  \vspace{2mm}
  \centering
  \includegraphics[width=0.48\textwidth]{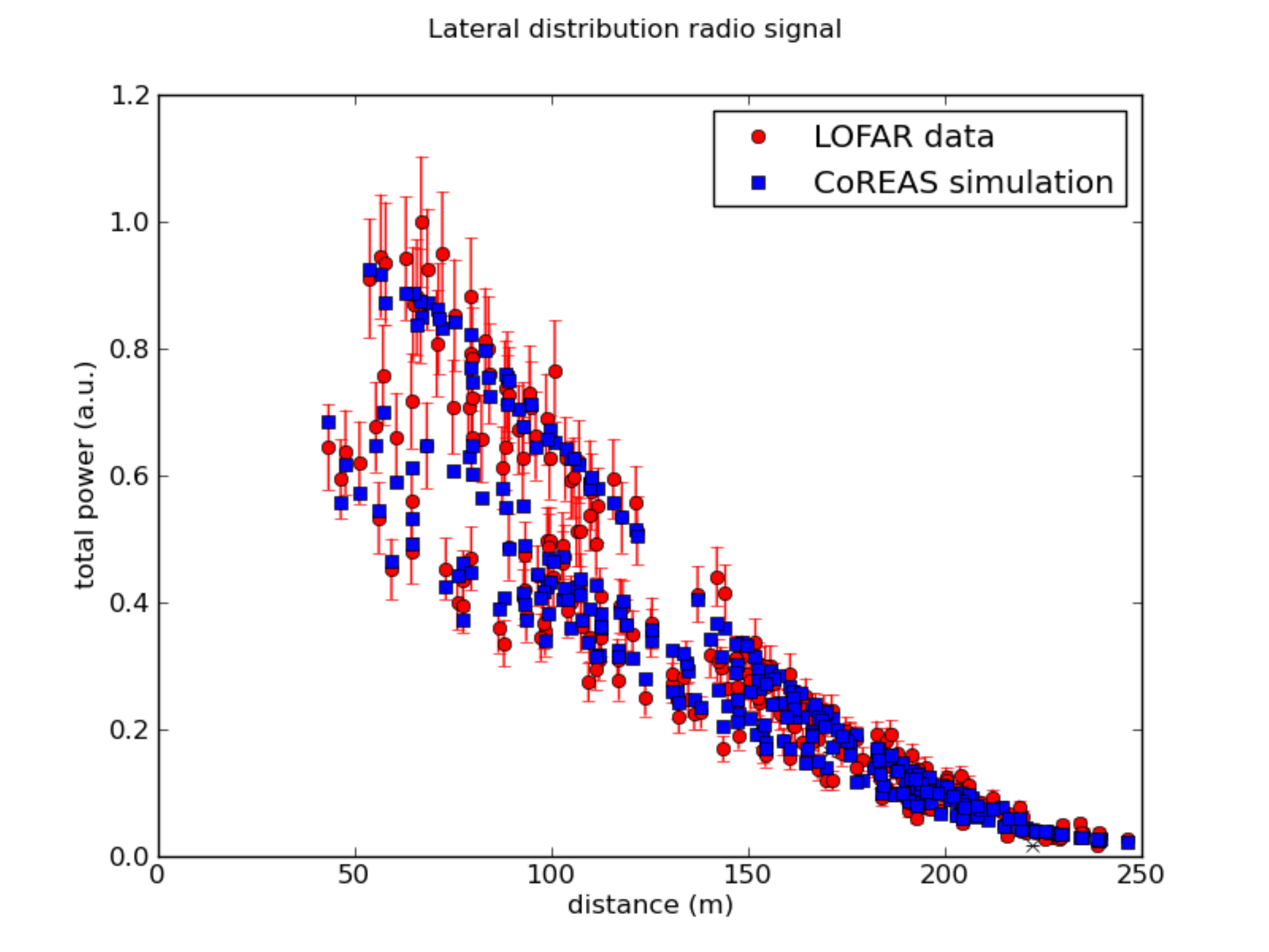}
  \caption{Comparison of a CoREAS simulation with a LOFAR event with a multitude of radio antennas with detected signal \citep{BuitinkLOFARIcrc2013}. The simulations can clearly reproduce the complexity of the measured radio lateral distribution. Please note that the absolute amplitude calibration has been fit with a universal scale factor and is thus not tested here.}
  \label{fig:lofarcomparison}
 \end{figure}

So, how well do we understand the radio emission? Models have been compared amongst each other, and while there are some significant differences in their predictions, these can usually be attributed to simplifications in the modelling. The two full-fledged microscopic Monte Carlo simulations CoREAS and ZHAireS, both of which do not make any particular assumption on the emission mechanisms but rather apply formalisms calculating the radio emission associated with the individual particle movement, agree well in their predictions \citep{HuegeARENA2012b}. Comparing them with data also shows good agreement, as is eximplified in Figs.\ \ref{fig:aeracomparison} and \ref{fig:lofarcomparison}. The shape of the measured lateral distribution is very well-reproduced, including asymmetries due to the superposition of geomagnetic and charge-excess emission as well as Cherenkov bumps. The currently most uncertain quantity is the absolute scale of the emission, which is difficult to determine experimentally and where some deviations have yet to be studied further \citep{SchroederLOPESCoREAS2013}.

\subsection{Triggering strategy}

A topic which has received significant attention in the past few years is the strategy for triggering radio detectors. In the pioneering phase of the modern radio detection experiments, information from particle detectors was used to trigger the radio read-out. This worked very well in particular for LOPES and CODALEMA. Once some experience had been gathered, methods for triggering on the radio signal itself were developed. Successful self-triggering was achieved with TREND \citep{Trend2011}, with a prototype setup at the Pierre Auger Observatory \citep{RevenuRAuger2012}, with the latest configuration of CODALEMA \citep{MachadoIcrc2013}, and with AERA \citep{HuegePisa2009}. There are, however, difficulties in particular with transient radio frequency interference which mimicks cosmic ray signals and often leads to either high trigger rates or high signal thresholds. If the high trigger rates can be maintained (which can be problematic due to bandwidth limitations of wide-spread detectors communicating via wireless links), it is possible to identify cosmic rays through an offline analysis usually involving a coincidence search with particle detectors. The radio data sets collected with such schemes to date, however, had a very low purity, and so far no convincing solution has been presented for a radio detector which works on its own and does not use information from other detectors to identify the cosmic ray events in the data set. One can, however, rightly ask whether self-triggered radio detection is actually an important goal. It has become increasingly clear in recent years that to solve the riddles of the origin of cosmic rays, as much information as possible has to be gathered on individual cosmic rays, especially at the highest energies. Hybrid detection is cleraly superior to detection with only one type of detector, and thus the general goal should anyway be to combine radio detection with other detection methods, which can then also help provide a trigger for radio measurements.

\subsection{Interferometric analysis}

\begin{figure*}[!htb]
  \vspace{2mm}
  \centering
  \includegraphics[width=0.4\textwidth]{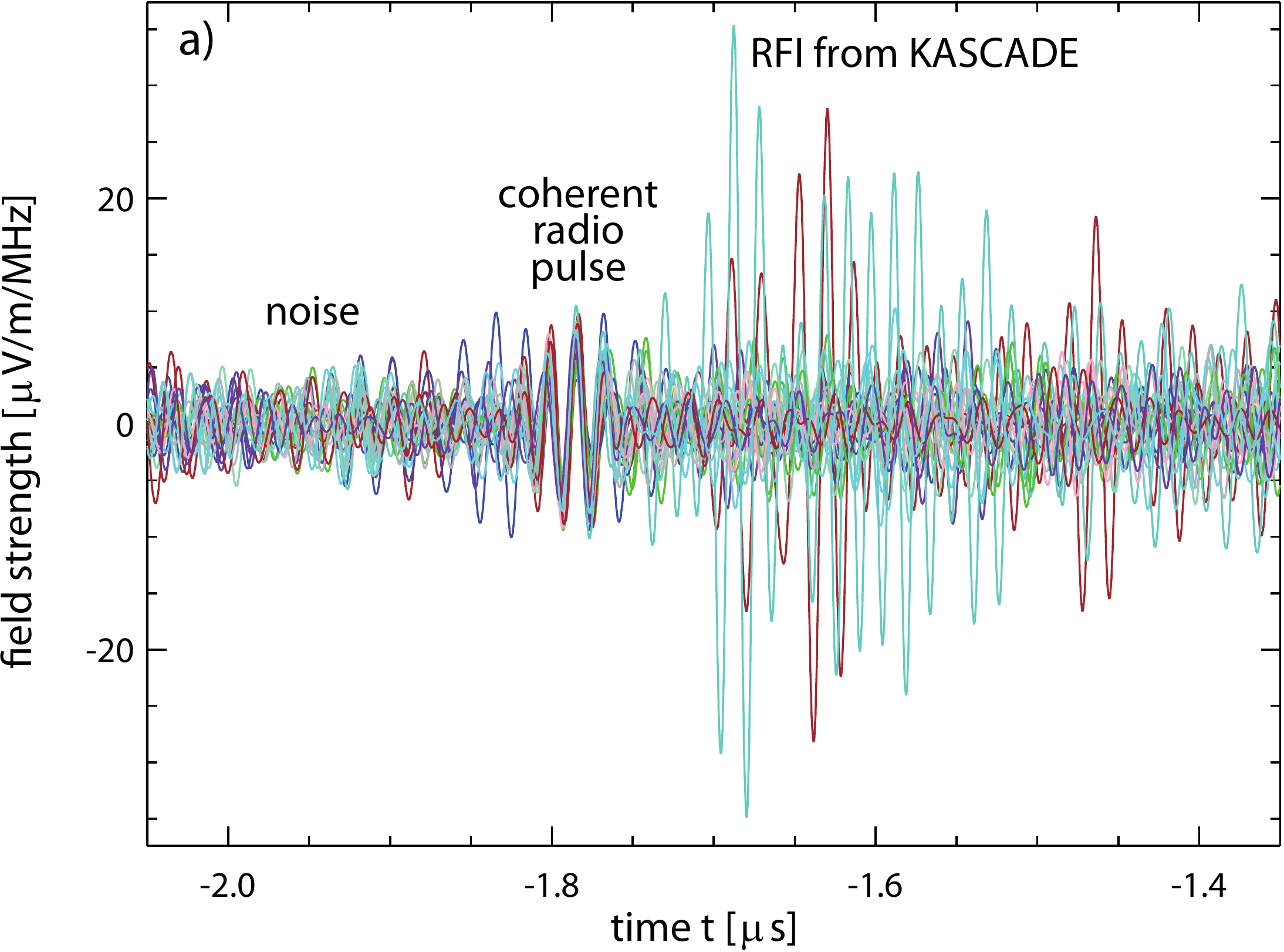}
  \includegraphics[width=0.4\textwidth]{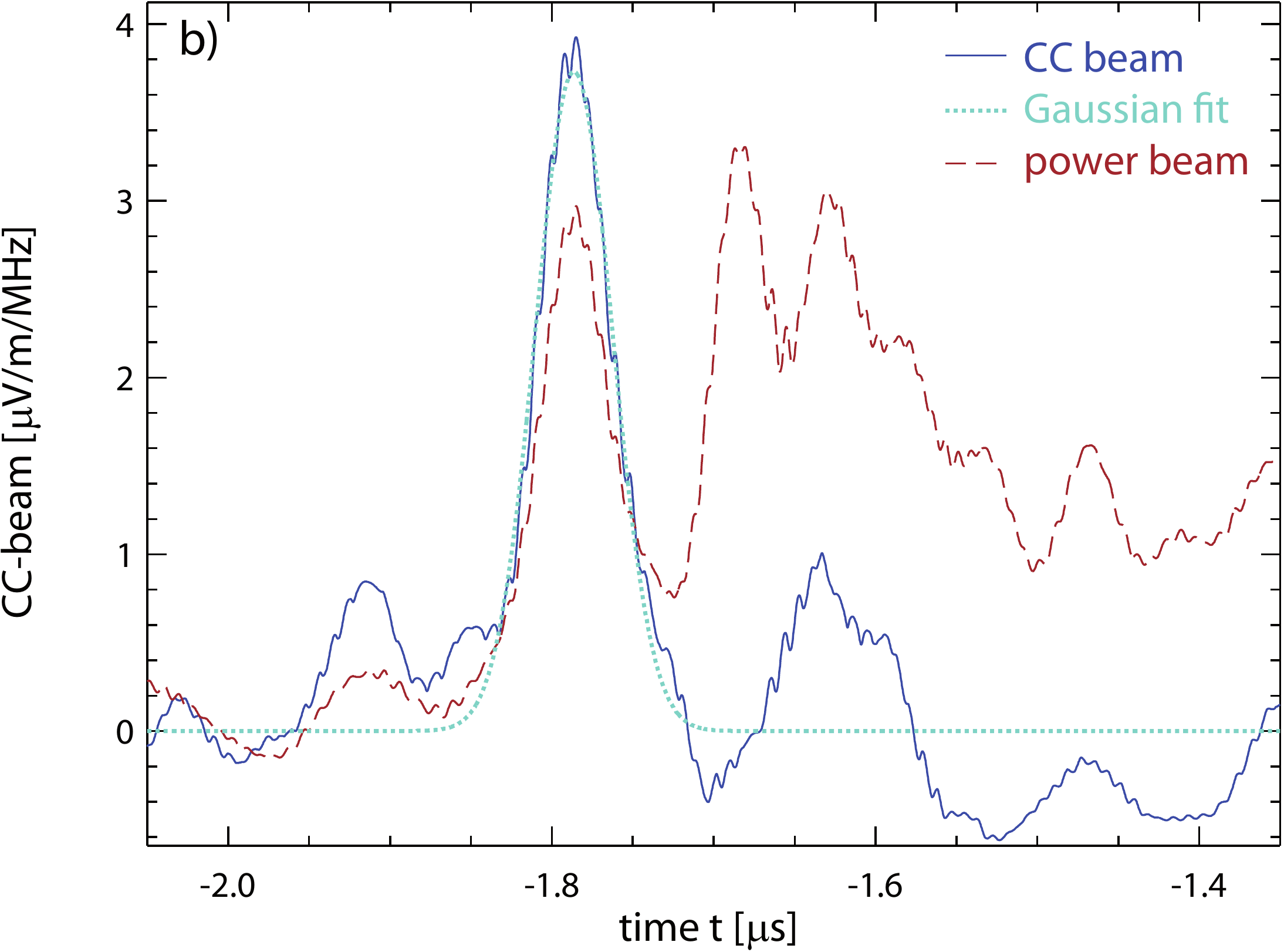}
  \caption{Electric field traces measured by individual LOPES antennas (left) and the cross-correlation averaged over all possible pairs of antennas \citep{SchroederLOPESCoREAS2013}. This interferometric analysis allows one to reliably detect radio signals with a signal-to-noise ratio way too small for detection and identification in individual antennas.}
  \label{fig:interferometry}
 \end{figure*}

Another topic related to the identification of radio events in the presence of noise is the way how radio data are analyzed. A straight-forward approach is to use techniques analogous to those used for analyzing particle detector data. An algorithm searches for pulses that reach a certain threshold over the noise level, tags the arrival times of the pulses in the antennas, and then fits a wavefront model to the arrival time distribution. This is the strategy that is currently followed by most of the radio detection experiments. It uses, however, only part of the available information. As the radio emission from air showers is coherent, one can also exploit the phase information of the radio signals. One way of doing this is by correlating the radio signals measured in different antennas with each other in an interferometric analysis similar to the ones used in radio astronomy. This technique has been applied successfully from the start in the LOPES experiment \citep{FalckeNature2005, SchroederLOPESCoREAS2013}. In fact, LOPES would never have been able to identify radio signals from air showers in its quasi-industrial environment withouth the use of interferometric techniques. This is illustrated in Fig.\ \ref{fig:interferometry}. On the left, the signal traces from individual LOPES antennas are shown. It is impossible to identify the actual cosmic ray radio signal by looking only at individual traces. However, the air shower radio signal, unlike the noise emanating from the KASCADE particle detectors and environmental noise sources, is coherent. A correlation analysis, shown on the right, can determine the coherent part of the signal (CC-beam), which clearly sticks out from the total power. Note that summing up the signals of the individual antennas is inferior to an actual interferometric correlation. Application of such interferometric techniques requires very good timing calibration of the detectors (a rule of thumb is to aim for 30$^{\circ}$ of phase at the highest used frequency, corresponding to $\sim 1$~ns in the case of upper frequency limits of $80$~MHz). Also, for larger radio detector arrays, the approaches developed so far will have to be extended to accomodate the fact that the radio pulses become broader with increasing lateral distance of the antennas.

\subsection{Direction resolution}
 
\begin{figure}[!htb]
  \vspace{2mm}
  \centering
  \includegraphics[width=0.45\textwidth]{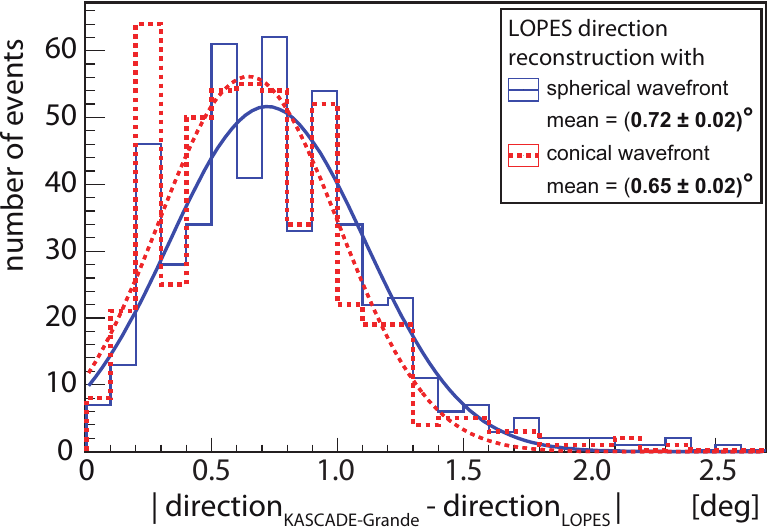}
  \caption{Illustration of the combined KASCADE-LOPES direction resolution achieved using various models for the form of the radio emission wavefront \citep{SchroederARENA2012}. Larger arrays are expected to have even better angular resolution.}
  \label{fig:direction}
 \end{figure}

Using interferometric techniques, the sky can be mapped and the arrival direction of the cosmic ray radio signal can be determined with very high precision. For instance, it has been demonstrated by the LOPES experiment that a combined precision of the arrival direction reconstruction between the particle detectors of KASCADE and the LOPES radio antennas of 0.65$^{\circ}$ could be reached \citep{SchroederARENA2012}, as is shown in Fig.\ \ref{fig:direction}. With larger arrays, an even higher angular resolution should be achievable. To achieve the highest possible angular resolution, it is important to use a proper model for the wavefront of the radio signal. As the source is not at infinity, the wavefront is not planar. For a long time, a spherical wavefront (corresponding to a point source) has been used, but recently it has become clear that a conical wavefront (corresponding line source) or even hyperbolical wavefront fits the data best \citep{SchroederARENA2012}.

\subsection{Energy determination}

\begin{figure*}[!htb]
  \vspace{2mm}
  \centering
  \includegraphics[width=0.4\textwidth]{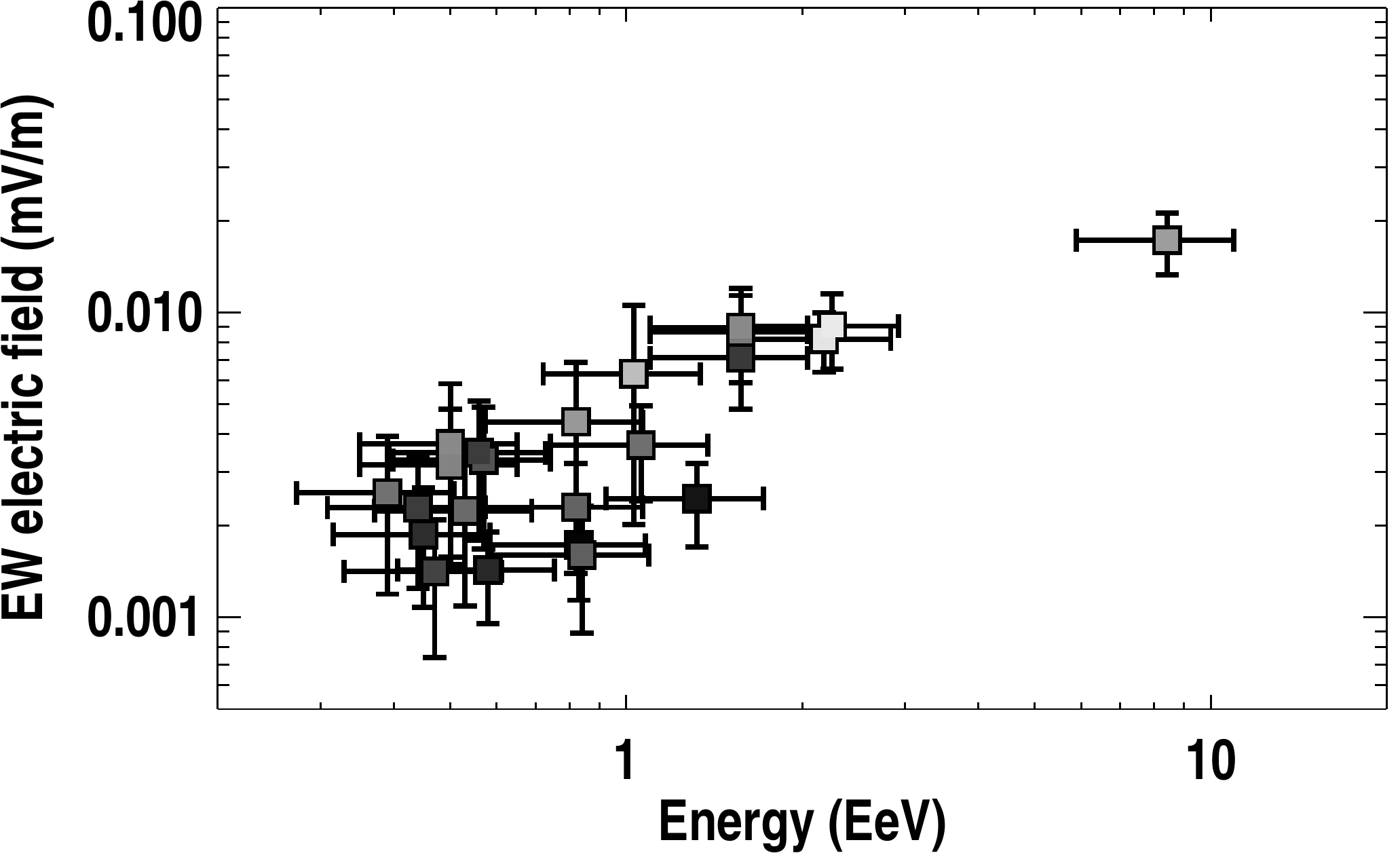}
  \hspace{2mm}
  \includegraphics[width=0.44\textwidth]{icrc2013-1294-17.pdf} \\
  \hspace{5mm}
  \includegraphics[width=0.38\textwidth]{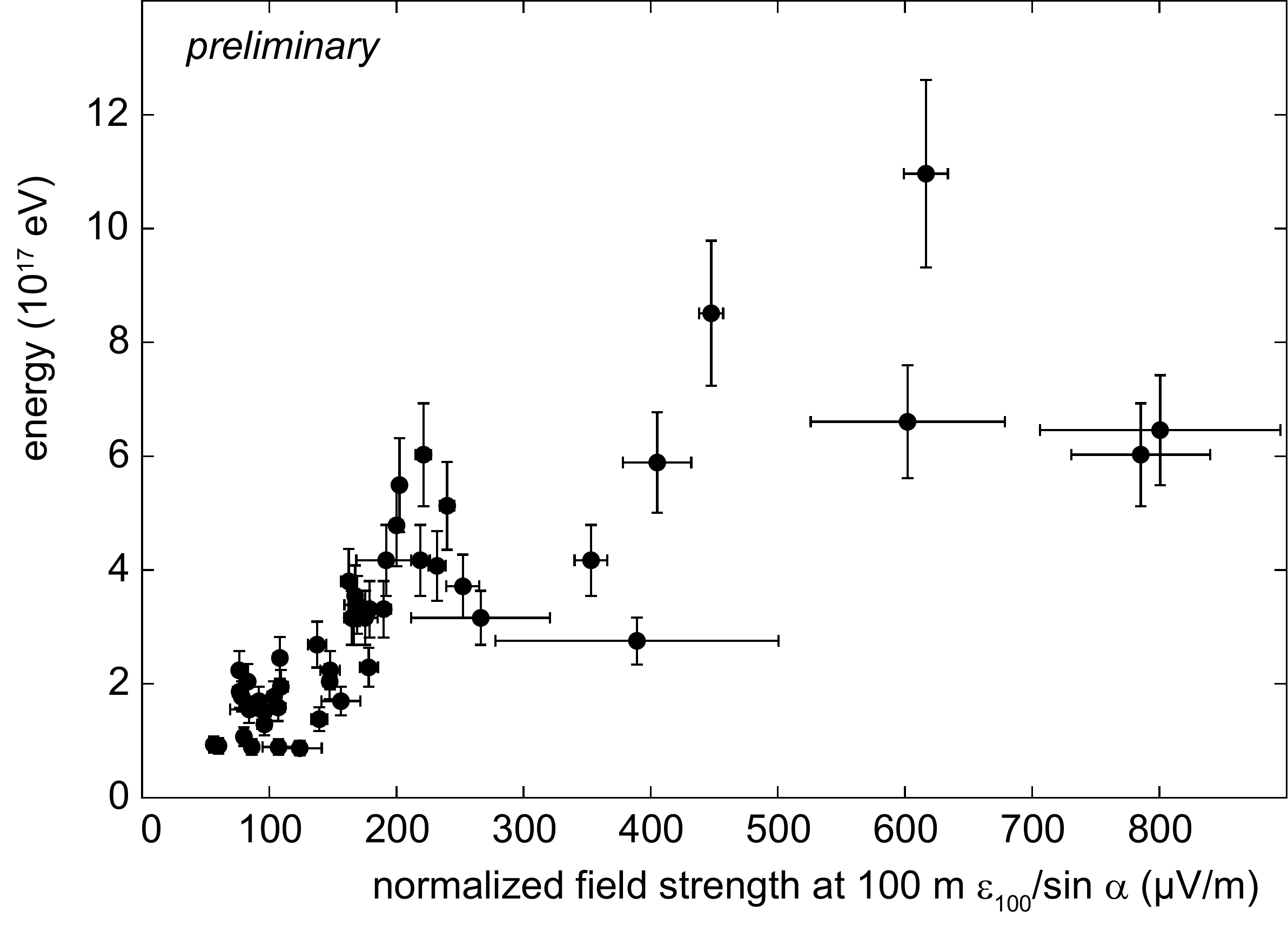}
  \hspace{2mm}
  \includegraphics[width=0.46\textwidth]{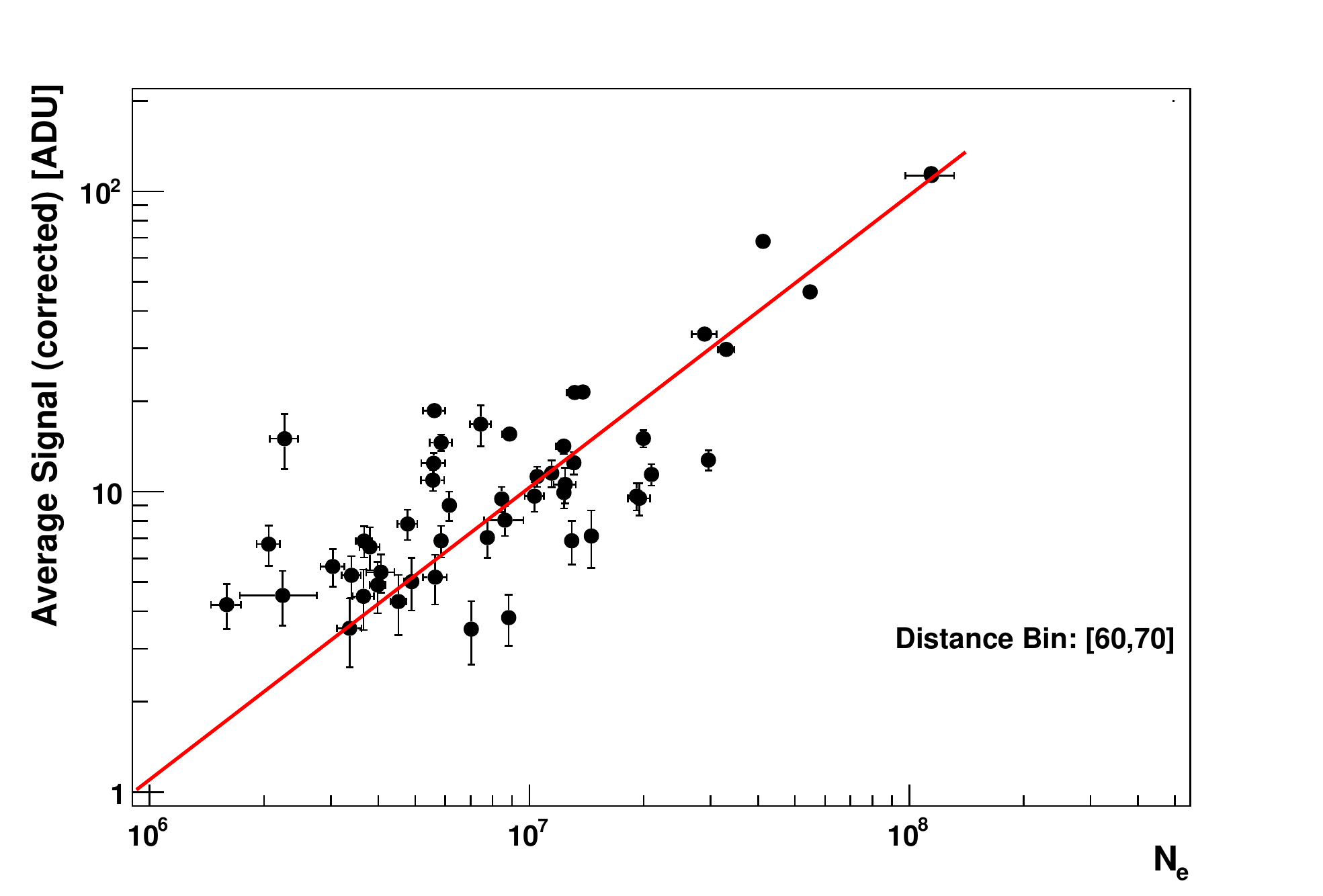}
  \caption{Clear correlations between the energy of the primary cosmic ray and the amplitudes of the measured radio pulses have been found by many experiments. Shown here are results from a prototype setup at the Pierre Auger Observatory (top left, \citep{RevenuRAuger2012}), the CODALEMA experiment (top right, \citep{RevenuARENA2012}), the Tunka-Rex experiment (bottom left, \citep{SchroederTRexIcrc2013}) and LOFAR (bottom right, \citep{NellesIcrc2013}.}
  \label{fig:energy}
 \end{figure*}

One of the most important quantities characterising a cosmic ray primary is its energy. Because the radio emission from air showers is coherent and because there is no significant attenuation of the signal in the atmosphere, the total electric field strength is basically proportional to the number of radiating particles, which in turn scales approximately linearly with the energy of the primary cosmic ray. This linear correlation between electric field amplitude and cosmic ray energy had already been observed in the 1970s. All modern experiments confirm this relation, as can be seen in Fig.\ \ref{fig:energy}. Note, however, that these results are mostly of a preliminary nature because experimental characteristics have not been completely unfolded, etc.

\begin{figure}[!htb]
  \vspace{2mm}
  \centering
  \includegraphics[width=0.48\textwidth]{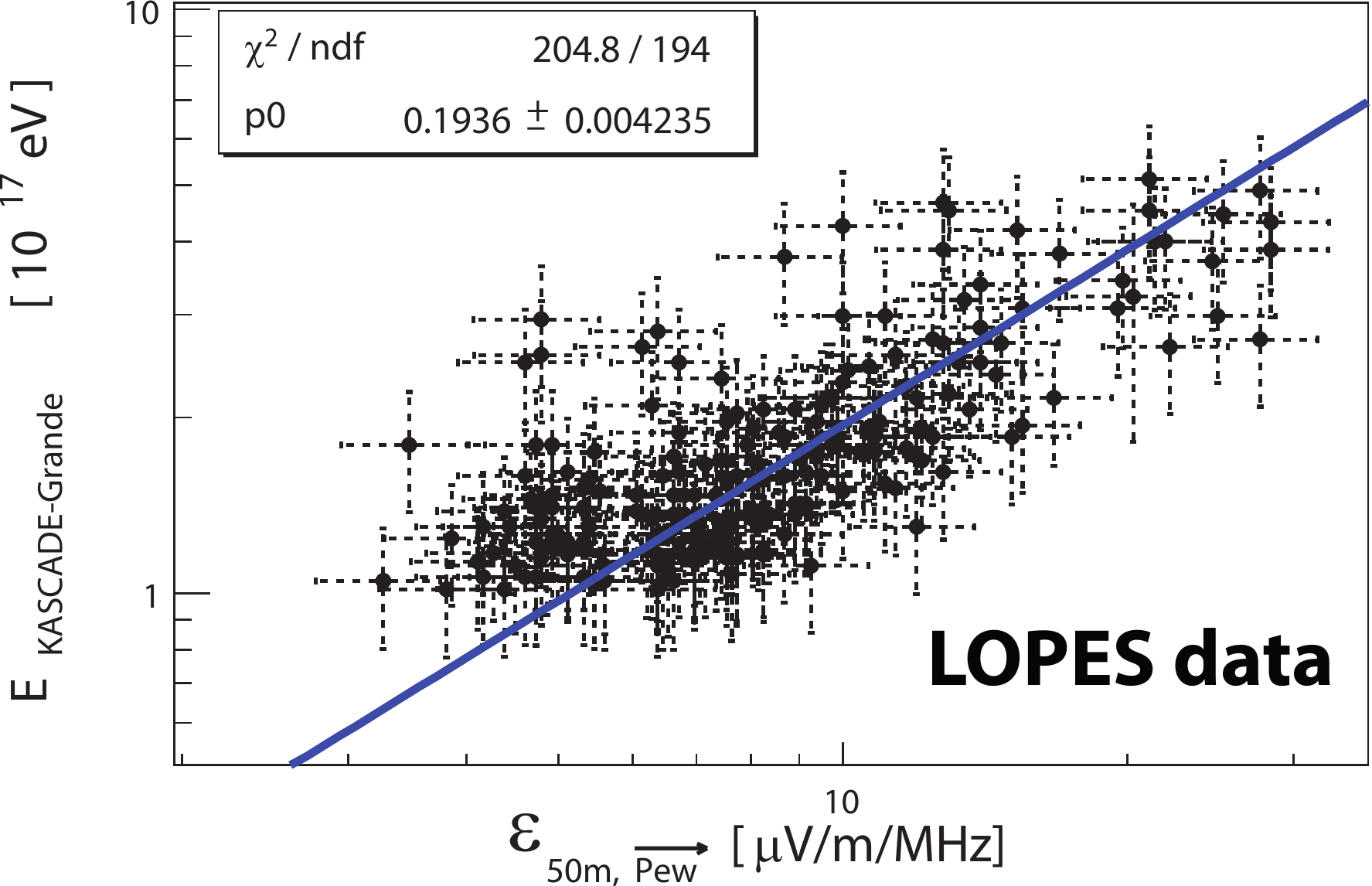}
  \caption{A linear correlation between the amplitude of the radio pulses measured by LOPES and the energy reconstructed by KASCADE is evident \citep{PalmieriIcrc2013}. The combined KASCADE-LOPES uncertainty is of order 20\%, and is probably dominated by the energy resolution of KASCADE.}
  \label{fig:lopesenergy}
 \end{figure}

A very detaild study has recently been presented by the LOPES Collaboration \citep{PalmieriIcrc2013}, one result of which is shown in Fig.\ \ref{fig:lopesenergy}. The study confirms that there is an optimum lateral distance at which the radio electric field amplitude is not affected by shower-to-shower fluctuations and can be used as a robust energy estimator. The electric field amplitude shows a linear correlation with the energy of the primary particle as determined with the KASCADE particle detectors. The combined uncertainty of KASCADE and LOPES is quantified as 20-30\% (depending on the shower zenith angle). Of that, the energy resolution of the KASCADE particle detectors alone is of order 20\%. Simulation studies predict that the precision of the radio energy determination could intrinsically be as good as 6-8\%. It is very hard to verify this, as the ``reference'' energy (in the case of LOPES provided by KASCADE) would need to be on the same level of precision.

An important fact to keep in mind is that radio emission directly probes the electromagnetic component of the air shower. Using radio detection to study the fraction of the energy in the electromagnetic cascade could thus be a means to study hadronic interaction models, by determining the fraction of energy which is not reflected in the radio signal as it goes into hadronic or muonic channels.

\subsection{Mass sensitivity}

The final parameter necessary for cosmic ray studies is the mass of the primary particle. This is usually estimated by measuring a mass sensitive parameter, the most common of which is the depth of shower maximum $X_{\mathrm{max}}$. In a simulation-based study \citep{HuegeUlrichEngel2008} it was already found in 2008 that the slope of the lateral distribution of the radio signal is correlated with $X_{\mathrm{max}}$. The reason for this is the relativistic beaming of the radio emission in the forward direction (as the source particles move approximately with the speed of light along the shower axis). The further the source is away (i.e., the smaller the value of $X_{\mathrm{max}}$), the larger is the illuminated area, or equivalently the flatter is the lateral distribution of the radio signal. This is a geometrical effect, and although the modelling efforts have made enormous progress since the original study, the results remained qualitatively valid and have been confirmed with various other models.\footnote{One should keep in mind that while grammage governs the physics of the shower evolution, geometrical distances govern the physics of the radio emission. In fact, the slope of the radio lateral distribution thus probes the geometrical distance of the source, not $X_{\mathrm{max}}$. Given an atmospheric model and a zenith angle, the two can of course be easily related with each other.}

\begin{figure*}[!htb]
  \vspace{2mm}
  \centering
  \includegraphics[width=0.5\textwidth]{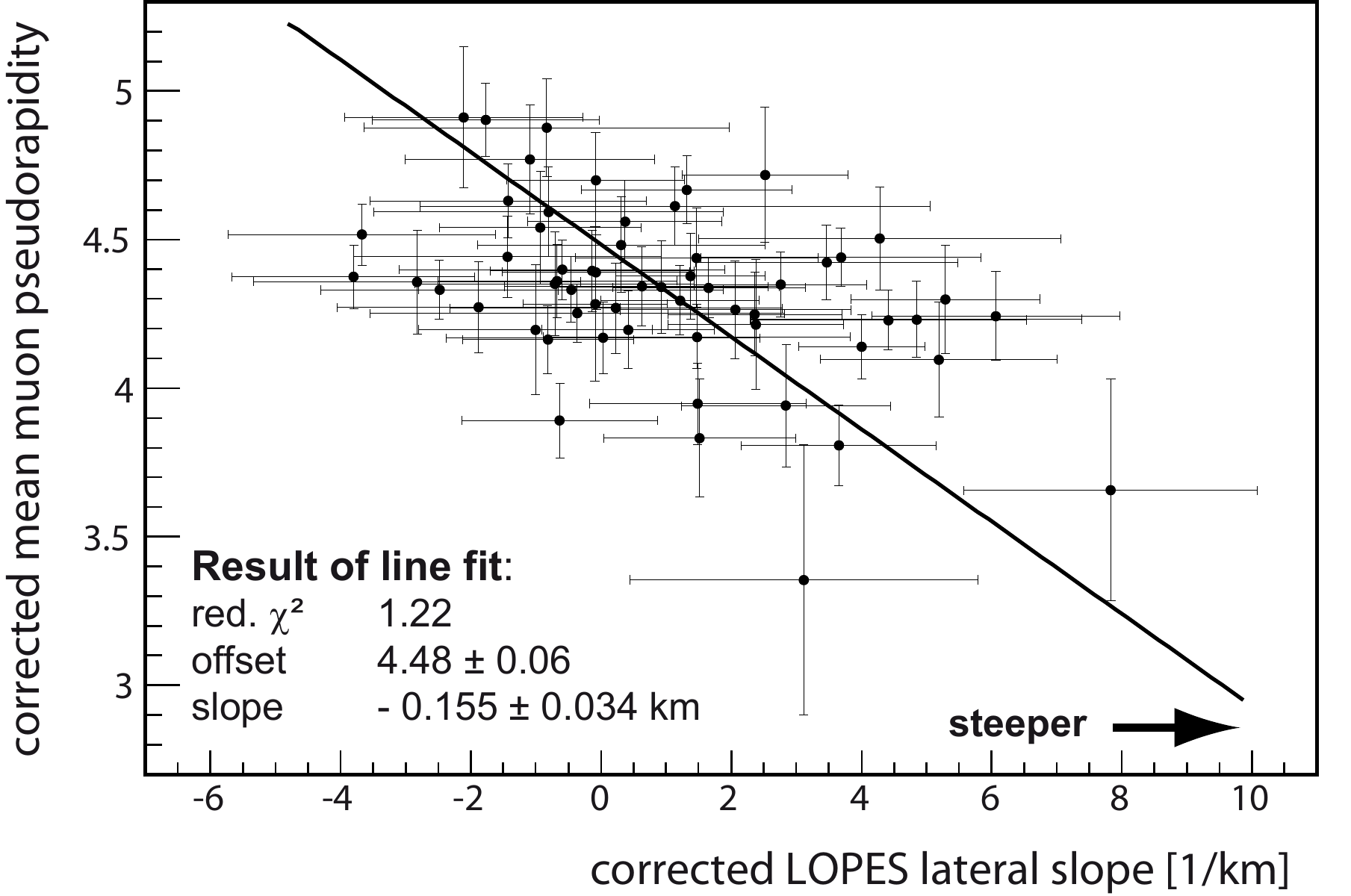}
  \includegraphics[width=0.41\textwidth]{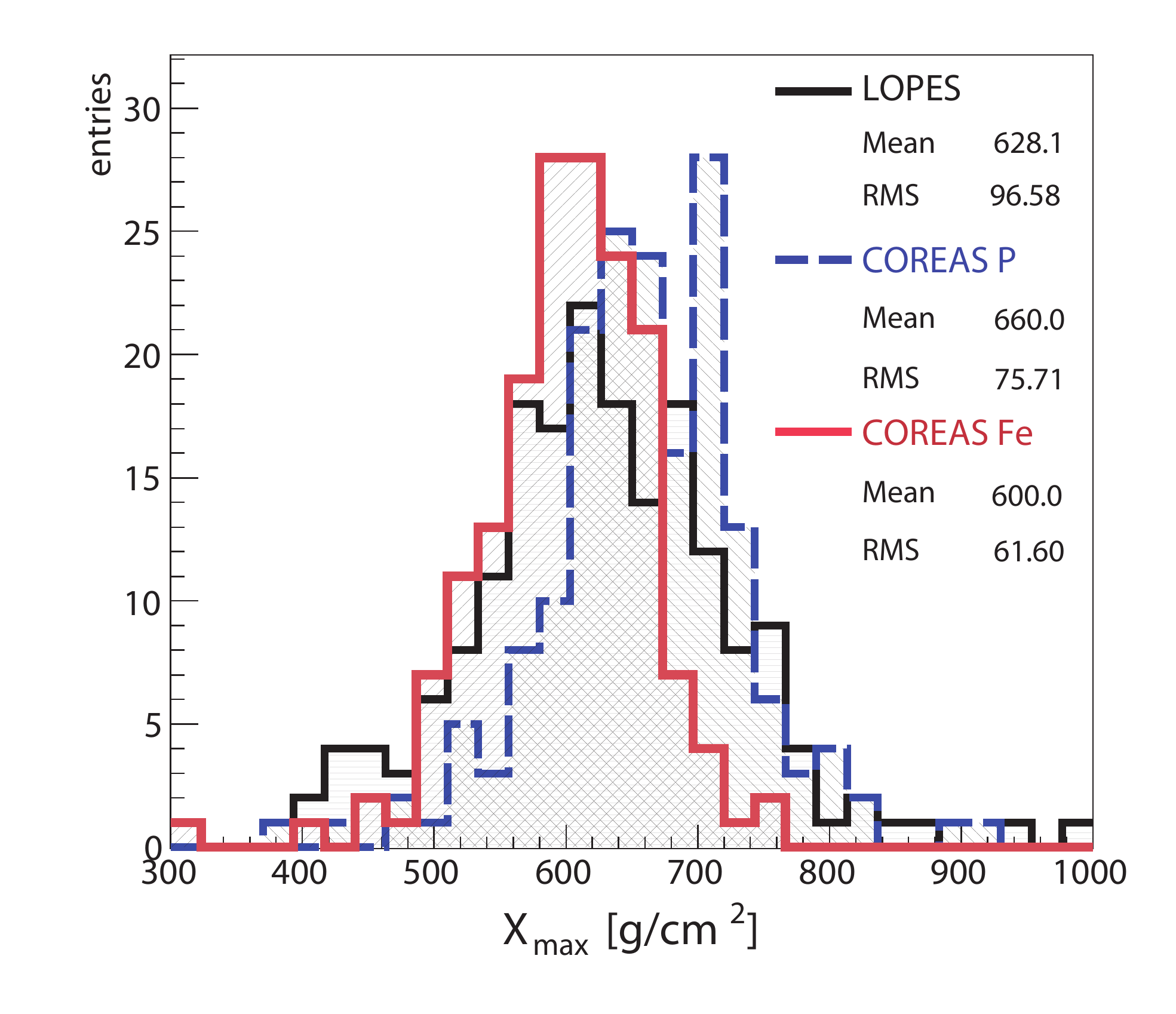}
  \caption{The mean muon pseudorapidity measured by KASCADE, a parameter associated with the production height of muons and thus with the longitudinal evolution of air showers, is correlated with the radio emission lateral slope as determined by LOPES (left) \citep{ApelArteagaBaehren2012c}. This is experimental evidence for the sensitivity of the radio signal to the air shower longitudinal evolution. Calibrating the radio lateral slope as a function of $X_{\mathrm{max}}$ using CoREAS simulations, the depth of shower maximum per event can be reconstructed from LOPES measurements (right) \citep{PalmieriIcrc2013}.}
  \label{fig:lopesmasses}
 \end{figure*}

There is by now experimental evidence that indeed the slope of the radio lateral distribution carries information on the longitudinal evolution of the air shower \citep{ApelArteagaBaehren2012c}. In Fig.\ \ref{fig:lopesmasses} it is shown that the slope of the radio lateral distribution measured by LOPES is correlated with the mean muon pseudorapidity, a quantity related to the muon production height, measured by KASCADE. This is a first confirmation of the expectations from simulations.

Using simulations (or later hybrid data), a ``calibration'' can then be made for the correlation between $X_{\mathrm{max}}$ and the radio lateral distribution slope. Once this is available, $X_{\mathrm{max}}$ can be determined from a radio measurement of the slope. The resulting distribution of $X_{\mathrm{max}}$ values as determined by the LOPES experiment on the basis of CoREAS simulations is shown in Fig.\ \ref{fig:lopesmasses}. The method developed so far has an intrinsic uncertainty of $\approx 50$~g~cm$^{-2}$ \citep{PalmieriIcrc2013}.

As was shown earlier (Fig.\ \ref{fig:footprint}), the radio lateral distribution is rather complex and thus carries a significant amount of information. If one has enough antennas to densely sample this distribution, one can not only use a 1-d approach as was done in the case of LOPES, but can use the full 2-d information for a global fit based on simulations. This was done in the case of LOFAR data and yielded an expected $X_{\mathrm{max}}$ resolution of $\approx 20$~g~cm$^{2}$ \citep{BuitinkLOFARIcrc2013}, which is comparable to the resolution of fluorescence detectors.

Of course, as of yet these results rest on simulations of the radio emission. One of the most important goals for the next few years will be to cross-check and cross-calibrate the $X_{\mathrm{max}}$ information determined with radio measurements with those measured by other techniques. This will be investigated in particular by AERA using the Auger fluorescene detectors and by Tunka-Rex using the Tunka Cherenkov light detectors.

In addition, other characteristics of the radio signal should be exploitable to gather information on the depth of shower maximum, in particular the opening angle of the radio wavefront \citep{SchroederIcrc2011} and the pulse shape or spectral slope of the frequency spectrum \citep{GrebeARENA2012}.

\section{Future directions}

\begin{figure*}[!htb]
  \vspace{2mm}
  \centering
  \includegraphics[width=0.34\textwidth]{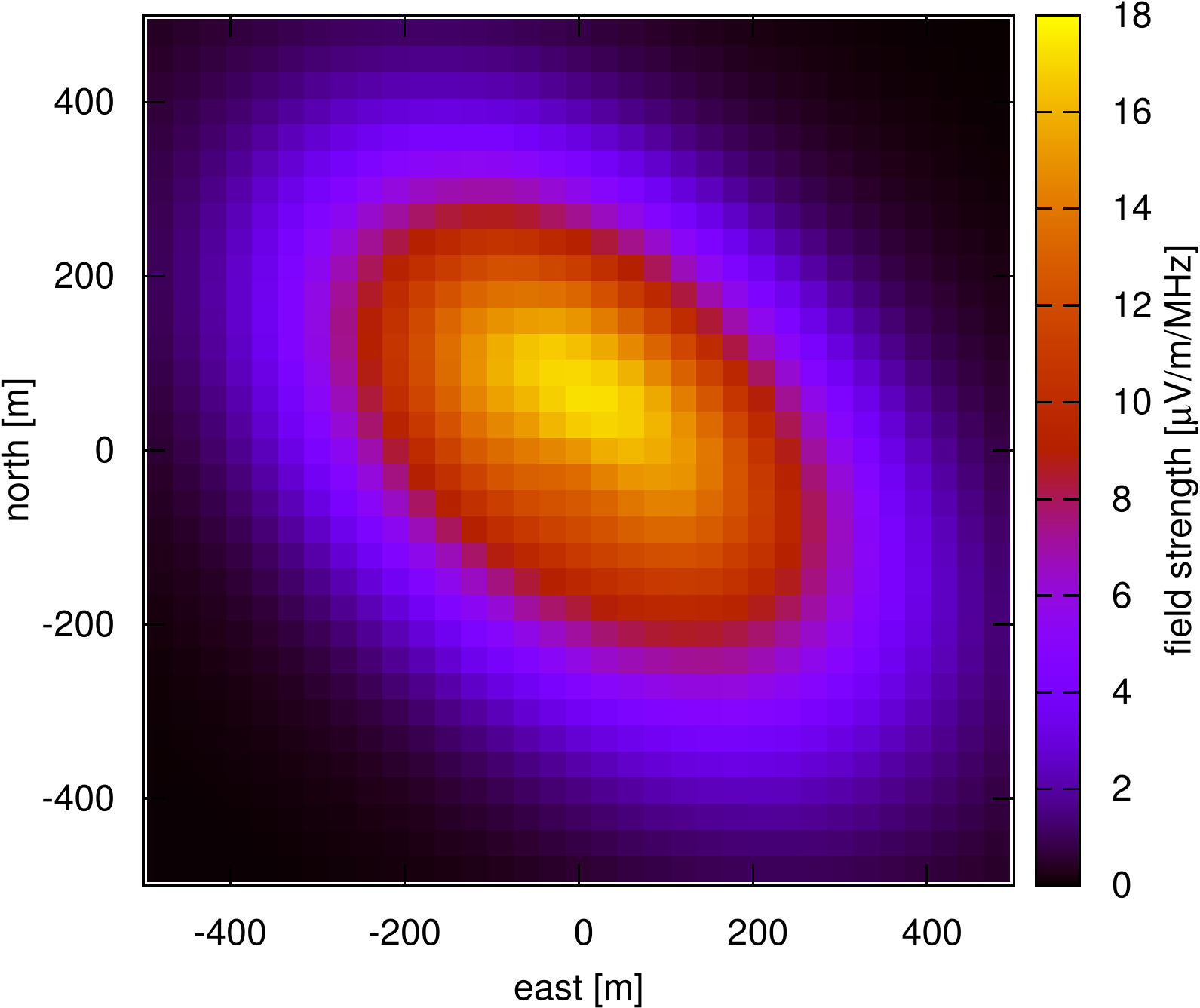}
  \hspace{0.05\textwidth}
  \includegraphics[width=0.34\textwidth]{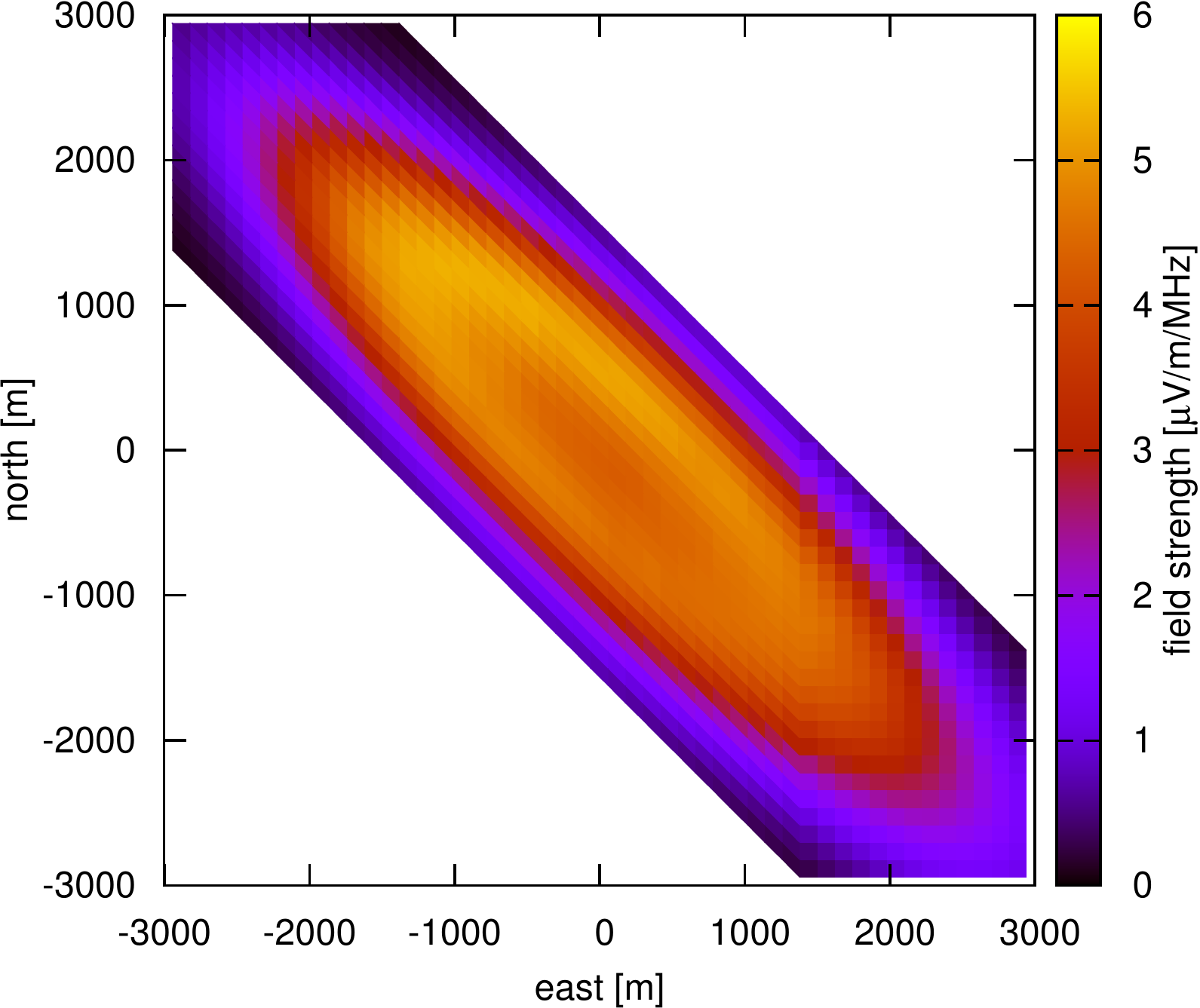}
  \caption{Simulated footprints of the radio signal for an air shower with a moderate zenith angle of 50$^{\circ}$ (left) and an inclined air shower with a zenith angle of 75$^{\circ}$ \citep{HuegeIcrc2013CoREAS}. The increase in the size of the footprint is dramatic and is caused by the receding of the radio source to greater geometrical distance from the antennas. For such inclined air showers, antennas could be spaced more than a kilometer apart and still allow coincident detection.}
  \label{fig:inclined}
 \end{figure*}

It is becoming more and more clear that the true power of radio detection lies in its hybrid application with other detectors, particular surface detector arrays. The use of a stand-alone radio detector seems much more difficult and less promising in comparison. The concepts which have been explored so far were very successful, but have also illustrated a major limitation: The maximum distance between radio antennas is limited to a few hundred meters. This is because the radio emission is strongly forward-beamed and thus, for near-vertical air showers, only illuminates a relatively small region on the ground \citep{HuegeIcrc2013CoREAS}. Increasing the energy does not remedy this problem, as the geometry is unchanged. In contrast, when going to very inclined air showers, the illuminated area becomes extremely large. Part of this is due to projection effects, which do not help to increase statistics. However, the more important effect is that for inclined air showers, the shower has to traverse more atmospheric matter, and in consequence the shower and thus the source of the radio emission recedes geometrically from the antennas. In other words, the emission, beamed into a similar angular opening angle, is now distributed over much larger areas. This is illustrated in Fig.\ \ref{fig:inclined}: the difference between 50$^{\circ}$ and 75$^{\circ}$ zenith angle is dramatic.

Efforts are being made to explore the detection of very inclined air showers, in particular by also measuring the vertical component of the electric field vector. This was pioneered with LOPES-3D \citep{ApelArteagaBaehren2012a} but is now also followed in other contexts because the vertical electric field component can become significant for very inclined air showers. An added benefit is that for inclined air showers, particle detectors measure the pure muonic component (the electromagnetic component has died out at ground level), whereas the radio detectors measure the pure electromagnetic component of the air shower. A combination of the two techniques could thus yield very good mass sensitivity.

\section{Conclusions}

Radio detection of cosmic rays has experienced an impressive renessaince, and has come a very long way in the past decade. Today, we master the detection of the radio signals, have achieved a detailed understanding of the emission physics, and have worked out how to reconstruct the direction and energy of the primary cosmic rays with very good precision. There are strong indications (from simulations and data) that it will also be possible to extract information on the mass of the primary particle from radio measurements with good precision. The methods for $X_{\mathrm{max}}$ determination based on simulations will be tested with experimental data in the coming few years, in particular by AERA and Tunka-Rex. Finally, while detection of near-vertical air showers turns out to be limited by a maximum distance between antennas of a few hundred meters, radio detection of very inclined air showers has the potential to work on extremely large scales and will thus be pursued specifically in the coming years.

\section*{Acknowledgements}

I would like to thank all colleagues in the field for fruitful discussions and collaboration, and the organizers of the ICRC for the honor to present this highlight contribution.


\end{document}